\documentclass[12pt]{article}
\usepackage{latexsym}
\usepackage{amsmath}
\usepackage{indentfirst}
\usepackage{cite} 

\long\def\ca#1\cb{} 

\newcommand{\ad}{^\dagger }

\newcommand{\AND}{\mbox{\small AND}}

\newcommand{\avg}[1]{\langle #1\rangle }
\newcommand{\becs}{\begin{cases}}
\newcommand{\bem}{\begin{matrix}}

\newcommand{\dya}[1]{|#1\rangle\langle#1|}

\newcommand{\encs}{\end{cases}}
\newcommand{\enm}{\end{matrix}}

\newcommand{\inpd}[2]{\langle#1|#2\rangle }
\newcommand{\ket}[1]{|#1\rangle }

\newcommand{\lra}{\leftrightarrow }

\newcommand{\mte}[2]{\langle#1|#2|#1\rangle }

\newcommand{\od}{\odot }

\newcommand{\ot}{\otimes }

\newcommand{\ra}{\rightarrow }
\newcommand{\Ra}{\Rightarrow }

\newcommand{\st}{\sqrt{2}}
\newcommand{\tm}{\times }

\newcommand{\vbb}{\mspace{1mu}|\mspace{1mu}}

\newcommand{\vbl}{\,\boldsymbol{|}\,}

\newcommand{\FC}{{\mathcal F}}

\newcommand{\HC}{{\mathcal H}}

\newcommand{\RC}{{\mathcal R}}

\newcommand{\rB}{\textbf{r}}

\newcommand{\dl}{\delta }

\newcommand{\lm}{\lambda }

\newcommand{\om}{\omega }

\def\outl#1{\par{\medskip\noindent\hspace*{0.1cm}\bf
      \mathversion{bold}#1\mathversion{normal}\smallskip} }
   \def\xa{} \def\xb{}  

 \def\outl#1{}\def\xa{}\def\xb{}

\ca
 \def\outl#1{\par{\medskip\noindent\hspace*{.5cm}\bf
      \mathversion{bold}#1\mathversion{normal}\smallskip} }
 \long\def\xa#1\xb{} 
  
\cb

\begin{document}

\title{Nonlocality Claims are Inconsistent with Hilbert Space Quantum 
Mechanics}
\author{Robert B. Griffiths\thanks{Electronic address: rgrif@cmu.edu}\\
Department of Physics\\
Carnegie Mellon University\\
Pittsburgh, PA 15213}

\date{Version of 3 March 2020}
\maketitle

\xa
\begin{abstract} 
  It is shown that when properly analyzed using principles consistent with the
  use of a Hilbert space to describe microscopic properties, quantum mechanics
  is a local theory: one system cannot influence another system with which it
  does not interact. Claims to the contrary based on quantum violations of Bell
  inequalities are shown to be incorrect. A specific example traces a violation
  of the CHSH Bell inequality in the case of a spin-3/2 particle to the
  noncommutation of certain quantum operators in a situation where
  (non)locality is not an issue. A consistent histories analysis of what
  quantum measurements measure, in terms of quantum properties, is used to
  identify the basic problem with derivations of Bell inequalities: the use of
  classical concepts (hidden variables) rather than a probabilistic structure
  appropriate to the quantum domain. A difficulty with the original
  Einstein-Podolsky-Rosen (EPR) argument for the incompleteness of quantum
  mechanics is the use of a counterfactual argument which is not valid if one
  assumes that Hilbert-space quantum mechanics is complete; locality is not an
  issue. The quantum correlations that violate Bell inequalities can be
  understood using local quantum common causes. Wavefunction collapse and
  Schr\"odinger steering are calculational procedures, not physical processes.
  A general Principle of Einstein Locality rules out nonlocal influences
  between noninteracting quantum systems. Some suggestions are made for changes
  in terminology that could clarify discussions of quantum foundations and be
  less confusing to students.
\end{abstract}
\xb
\xa
\newpage

\tableofcontents
\xb
\section{Introduction \label{sct1}}
\xa
\xb
\outl{Qm nonlocality notion widespread. Bell $\leq$s.; Local realism; steering}
\xa

The notion is widespread in popular articles, but also in many technical
papers, review articles, and books, that quantum mechanics is `nonlocal', in
some way that contrasts with the locality of classical physics. Here are a few
almost random selections from this vast literature:
\cite{AlGl09,HnSh18,QnVB14,Dng18,Brao14,Mdln11c,Nrsn17}.
In particular, quantum mechanics is, we are told, inconsistent with `local
realism' \cite{Hnao15ea,Wsmn15} because it predicts, and numerous experiments
confirm, a violation of Bell inequalities, and this means that if the
quantum-mechanical world is real there exist nonlocal influences which act
instantaneously over arbitrarily large distances. And if two distant systems
are in a suitable entangled quantum state, a measurement on one of them can
instantaneously influence the other through a process known as `steering'
\cite{Schr36,WsJD07,CvSk17,Tsao18}.

\xb
\outl{Maudlin-Werner debate. Black box approach}
\xa

To be sure, such claims have not gone unchallenged. Notable among more recent
discussions is an interchange between a proponent of nonlocality, Tim Maudlin
\cite{Mdln14,Mdln14b} and an advocate of quantum locality, Reinhard Werner
\cite{Wrnr14,Wrnr14b} that appeared in a special issue of the Journal of
Physics A published on the fiftieth anniversary of a famous paper \cite{Bll64b}
by John Bell; there was also a follow-up preprint \cite{Wrnr14c} by Werner. It
is of interest that neither protagonist in this debate actually applied quantum
theory to properties and processes taking place in a microscopic quantum
system. Instead, both used what might be called a `black box' approach: A
macroscopic preparation of the quantum system is followed later by a
measurement with a macroscopic output (`pointer position' in the antique but
picturesque terminology of quantum foundations), with the discussion based upon
quantum predictions of the relationship of input and output, without reference
to what might be going on at the microscopic quantum level at an intermediate
time. In Maudlin's case no reference to such goings on was needed for his
arguments, whereas Werner employed an operational approach to quantum theory in
which microscopic concepts are deliberately omitted. While a black box approach
can sometimes be useful in this as in other areas of science, the claim
of the present paper is that the locality issue is best addressed by opening
the black box and examining what happens inside it, using consistent quantum
principles. In particular it is important to understand how quantum
measurements can reveal microscopic quantum properties; something often
assumed by experimenters who design and build apparatus, but not properly
discussed in introductory (or advanced) quantum textbooks.

\xb
\outl{Wavefunction collapse: a tool to obtain conditional probabilities}
\xa

One source of the nonlocality idea is the widespread belief that measurements
`collapse' quantum wave functions. If one of two (or more) separated quantum
systems described by an entangled wavefunction is measured, then it is indeed
possible to discuss the post-measurement situation using a `collapsed' wave
function, and this no doubt contributes to the belief that there must be
nonlocal influences in the quantum world. However, in this situation the
wavefunction is merely a convenient tool for obtaining certain conditional
probabilities that can be calculated by other methods that do not suggest any
sort of nonlocal influence, as explained in Sec.~\ref{sct6}.

\xb
\outl{No-signaling $\Ra$ experiments cannot detect nonlocal influences}
\xa

To be sure, those who claim that instantaneous nonlocal influences are present
in the quantum world will generally admit that they cannot be used to transmit
information; this is known as the `no-signaling' principle, widely assumed in
quantum information theory. This means that such influences (including
wavefunction collapse) cannot be directly detected in any experiment. The
simplest explanation for their lack of influence is that such influences do not
exist.

\xb
\outl{Correlations in Cl systems $\lra$ provide an analogy for a Qm common
  cause}
\xa

In classical physics two systems far apart can exhibit statistical correlations
that allow a measurement on one of them to reveal some property of the other.
No instantaneous nonlocal influences need be invoked if such correlations
result from a local common cause at some time in the past. As explained in
Sec.~\ref{sbct5.3}, an analogous kind of \emph{quantum} common cause can be
used to understand quantum correlations that violate Bell inequalities, thus
removing any need for nonlocal influences.

\xb
\outl{Overview of paper. Sec.~\ref{sct2}. CHSH violated where situation is purely
  local. }
\xa

The arguments that support these conclusions are carried out in several steps.
First, in Sec.~\ref{sct2} the CHSH Bell inequality \cite{CHSH69} is shown to be
violated in a purely local situation where nonlocality plays no role. The key
point is that the CHSH inequality employs \emph{classical} hidden variables in a
situation where a proper \emph{quantum} description requires the use of
\emph{noncommuting} operators to represent physical quantities. That classical
physics fails in the quantum domain is not at all surprising. What is
surprising is that this fact \cite{Fne82b} has been overlooked or
ignored in much of the literature that claims quantum mechanics is nonlocal.

\xb
\outl{Sec.~\ref{sct3}: What projective Qm measurements measure}
\xa

\xb
\outl{Qm particles not simultaneously in two different locations}
\xa

\xb
\outl{Sec.~\ref{sct4}: Bell $\leq$ derivation factorization condition assumes 
Cl HVs}
\xa

Next, Sec.~\ref{sct3} is devoted to an elementary discussion of projective
quantum measurements and what they reveal about properties of measured systems
\emph{before} a measurement takes place. This is an essential part of opening
the black box, and fills in a serious lacuna in textbooks. It justifies the
belief of many experimental physicists that the apparatus they have carefully
constructed and tested actually measures what it was designed to measure. A
proper understanding of measurements disposes of another type of supposed
quantum nonlocality: that quantum particles can simultaneously be in two
different locations.
The tools that allow quantum measurements to be understood in a rational manner
are then applied in Sec.~\ref{sct4} to a serious defect that enters many if not
all derivations of Bell inequalities: a key factorization assumption that is
supposed to represent the absence of nonlocal effects employs \emph{classical}
hidden variables that are inconsistent with Hilbert space quantum theory.

\xb
\outl{Sec.~\ref{sct5}: EPRB contains false counterfactual assumption }
\xa

\xb
\outl{EPRB correlations explained in terms of \emph{Qm} common causes}
\xa

\xb
\outl{Alice can infer properties of, but not influence or ``steer'' Bob's
  particle}
\xa

The much-discussed Einstein-Podolsky-Rosen (EPR) argument is examined in
Sec.~\ref{sct5}, beginning in Sec.~\ref{sbct5.1} with Bohm's formulation in
terms of two spin-half particles. If one assumes, contrary to EPR, that Hilbert
space quantum mechanics is \emph{complete}, this undermines a key
counterfactual assumption about quantum measurements implicit in their work, an
assumption that has nothing to do with locality. Following this, in
Sec.~\ref{sbct5.3} it is shown that the experimentally observed correlations
which violate Bell inequalities can be understood as arising from local
\emph{quantum} common causes, and hence in no need of explanations based upon
instantaneous nonlocal influences. An analogy from the classical world helps
understand why Alice's measurement of one of a pair of spin-half articles has
not the slightest influence on the other particle, located far away in Bob's
possession, though she is able to infer something about its properties. In no
sense can she control or influence or `steer' Bob's particle.

\xb
\outl{Sec.~\ref{sct6} Wavefn collapse, Einstein locality}
\xa

In Sec.~\ref{sct6} it is argued that wavefunction collapse, while it can be
used to calculate correlations, is simply a mathematical tool, and should
\emph{not} be understood as a nonlocal physical process. Indeed, a quite
general \emph{Principle of Einstein Locality} states that noninteracting
systems cannot influence each other, whether or not they are in an entangled
state. So in this respect discussions, as in \cite{CvSk17,Tsao18}, of
Schr\"odinger steering are misleading.

\xb
\outl{Sec.~\ref{sct7}: Summary + suggestions for terminology}
\xa

A summary of the results of the paper are given in Sec.~\ref{sbct7.1}. This is
followed in Sec.~\ref{sbct7.2} with suggestions for changes in terminology
which might help clear up the confusion associated with long-standing, but
unsupportable, claims of quantum nonlocality, thus making quantum theory less
of an ordeal for students, and allowing more rapid progress in the study of
quantum foundations.

\xb
\outl{Earlier RBG work included, extended in this paper. CH opens black
  box}
\xa
 
The present paper incorporates, but also extends, material from some of the
author's earlier publications
\cite{Grff11,Grff11b,Grff15,Grff17b}.
The aim is to present a unified and comprehensive critique of quantum
nonlocality claims, based in large part on a consistent analysis of quantum
measurements. In order to understand in physical terms what is going on in the
quantum world, measurements themselves must be described as physical processes
governed by general quantum principles that apply to all processes. In
particular, macroscopic measurement outcomes must be connected with the prior
\emph{quantum} properties, represented by Hilbert subspaces (as in Sec.~III.5
of \cite{vNmn32b}), the apparatus was designed to reveal. The consistent
histories (CH) approach%
\footnote{For an overview of consistent histories see \cite{Grff14b}; a
  detailed treatment will be found in \cite{Grff02c}. The material in
  \cite{Grff17b} is of particular relevance to the present article.} %
provides the precise rules needed to do this, and is the foundation of the
discussions in Sections~\ref{sct3}, \ref{sct4}, \ref{sct5}, and \ref{sct6}.

\xb
\subsection{Notation and Abbreviations \label{sbct1.1}}
\xa

A quantum \emph{physical property} is represented by a Hilbert subspace
or its projector, as distinct from a \emph{physical variable} represented by a
Hermitian operator; see Sec.~\ref{sbct3.1}.

For the most part standard Dirac notation is employed, with the addition
that $[\psi]=\dya{\psi}$ denotes the projector onto a normalized pure state
$\ket{\psi}$.

CH is an abbreviation for the \emph{Consistent (or Decoherent) Histories}
formulation or interpretation of quantum mechanics.

CHSH is an abbreviation for the Bell inequality of 
Clauser, Horne, Shimony, and Holt \cite{CHSH69}.

EPR stands for Einstein, Podolsky, and Rosen and their paper [35].

PDI stands for \emph{projective decomposition of the identity}, see
\eqref{eqn9}.  

\xb
\section{Bell Inequalities \label{sct2}}
\xa

\xb
\outl{3 steps to nonlocality using Bell inequality violations}
\xa

A common route to the belief that the world is
nonlocal comes from the following sort of reasoning:
\begin{description}
\item[B1.] Bell (and others) derived inequalities involving correlations of
  separated quantum systems, inequalities which will always be satisfied if a
  certain locality condition (local causality or local realism) is satisfied .
\item[B2.] Starting with the work of Freedman and Clauser \cite{FrCl72},
  numerous experiments, among them \cite{AsGR81,Hnao15ea,Shao15ea,Gsao15ea},
  have shown, with ever increasing precision and control for errors and
  experimental loopholes, that experimentally measured correlations agree with
  the predictions of quantum mechanics and violate Bell inequalities.
\item[B3.] Therefore quantum mechanics, and the world it describes,
  must be nonlocal.
\end{description}

\subsection{The CHSH Inequality \label{sbct2.1}}
\xb
\outl{Formula $S=A_0B_0 +\cdots$}
\xa

To see what is wrong with this argument, consider the CHSH inequality
\cite{CHSH69}, one of the simplest Bell inequalities. It involves a
quantity
\begin{equation}
 S = A_0 B_0 + A_0 B_1 + A_1 B_0 - A_1 B_1 = (A_0+A_1)B_0 +(A_0-A_1)B_1
\label{eqn1}
\end{equation}
where the $A_j$ and $B_k$ on the right hand side are either classical random
variables taking the values $+1$ and $-1$, or quantum observables (Hermitian
operators) whose eigenvalues are $+1$ and $-1$, subject to the condition that
each $A_j$ commutes with every $B_k$. 

\xb
\outl{Cl case: $-2\leq S\leq 2$ $\Ra$ $|\avg{S}|\leq 2$}
\xa

In the classical case it is easy to see that because either $A_0+A_1$ or
$A_0-A_1$ must be zero, $S$ will lie between the limits
\begin{equation}
 -2 \leq S \leq 2,
\label{eqn2}
\end{equation}
so its average $\avg{S}$ must fall in the same interval, whence the CHSH
inequality:
\begin{equation}
 |\avg{S}| \leq 2.
\label{eqn3}
\end{equation}

\xb
\outl{Qm case: operators don't commute. $S$ can have eval as large as $2\st$}
\xa

By contrast, if the $A_j$ and $B_k$ are quantum Hermitian operators with
eigenvalues $\pm 1$ subject to the requirement that $[A_j,B_k]=0$ for every $j$
and $k$, it is easy to construct an example, see below, in which $S$ has
eigenvalues of $\pm2\st,0,0$, and thus using the eigenstate for the largest
eigenvalue to compute the average of $S$ will yield $\avg{S}=2\st$, in obvious
violation of the inequality \eqref{eqn3}. A key feature of this example is that
$A_0$ does not commute $A_1$, nor $B_0$ with $B_1$, and none of the four
summands in \eqref{eqn1} commute with the other three. There is no reason to
expect the eigenvalues of a sum of noncommuting operators to bear any simple
relationship with those of the summands, so the violation of \eqref{eqn3} in
the quantum case is not surprising. Nonlocality is irrelevant, as is
shown by the following example.

\subsection{Neon \label{sbct2.2}}


\xb
\outl{$^{21}$Ne spin 3/2. Let $\ket{00},\,\ket{01},\,\ket{10},\,\ket{11}$ be
  any orbasis}
\xa

The $^{21}$Ne nucleus has a spin of $3/2$, which is also the spin of a neutral
neon atom of this isotope; it has a low but nonzero natural abundance. Thus
the ground state of a $^{21}$Ne atom is 4-fold degenerate, and its quantum
mechanical description uses a four-dimensional Hilbert space $\HC$. Choose any
orthonormal basis for this space, and let the basis vectors carry binary
labels, thus $\ket{00},\,\ket{01},\,\ket{10},\,\ket{11}$. These could, for
example, be states in which the $z$ component of angular momentum $S_z$ for
some (arbitrary) direction $z$ takes on the values $+3/2,\,+1/2,\,-1/2,\,-3/2$
in units of $\hbar$, but any other choice would be equally good. 

\xb
\outl{Ground state $\HC=\HC_a\ot \HC_b$ with operators $A_j$, $B_k$,
  $M_{jk}=A_jB_k$ }
\xa

Next, as a matter of convenience, write $\HC$ as a tensor product
$\HC_a\ot\HC_b$ of two $2$-dimensional spaces with orthonormal bases
$\ket{0}_a$, $\ket{1}_a$, and $\ket{0}_b$, $\ket{1}_b$, respectively, related
to the previously chosen basis of $\HC$ through
\begin{equation}
 \ket{j,k} = \ket{j}_a\ot\ket{k}_b,
\label{eqn4}
\end{equation}
Finally, using this tensor product structure, define four operators
\begin{equation}
 A_0 = Z\ot I,\quad A_1 = X\ot I,\quad B_0= I\ot X,\quad B_1 = I \ot Z,
\label{eqn5}
\end{equation}
where $I$ is a $2\tm2$ identify matrix, while  $X$, and $Z$ are the Pauli
$x$ and $z$ matrices. Define
the products $M_{jk} = A_j B_k$:
\begin{equation}
 M_{00} =Z\ot X,\quad M_{01} =Z\ot Z,\quad 
 M_{10} =X\ot X,\quad M_{11} =X\ot Z, 
\label{eqn6}
\end{equation}
(where the subscripts label different operators, not matrix elements),
and a quantum version of \eqref{eqn1} takes the form:
\begin{equation}
 S = M_{00} + M_{01} + M_{10} - M_{11}.
\label{eqn7}
\end{equation}
Each $M_{jk}$ has eigenvalues $+1$ and $-1$, both doubly degenerate, and 
each does not commute with any of the other three, even though each $A_j$
commutes with each $B_k$. 

\xb
\outl{Measure $M_{jk}$ in $^{21}$Ne beam in 4 experiments $\ra$
$S=2\st$, violates CHSH}
\xa

Now imagine that a skilled experimenter is able to produce a beam consisting of
neon atoms of this isotope, each in the same (pure) hyperfine state
$\ket{\psi}$, and then, using a large number of runs, measures each $M_{jk}$
and finds its average value. Note that four separate runs of the experiment are
needed, since each $M_{jk}$ does not commute with the others. Adding up the
averages provides the average of $S$. Since the eigenvalues of the operator in
\eqref{eqn7} are $\pm 2\st,0,0$, if $\ket{\psi}$ is the eigenstate with the
largest eigenvalue, the average $\avg{S}$ of $S$ will be $2\st$, well outside
the range \eqref{eqn3}.

\xb
\outl{Experiment is local; violation of CHSH $\lra$ noncommutation}
\xa

In the case of this hypothetical $^{21}$Ne experiment all the atoms belong to a
single beam, and while the four different measurements require different
apparatus settings, they can all be carried out in the same physical location
in the same laboratory. Thus the violation of the CHSH inequality in this case
has nothing to do with nonlocality. Instead it has everything to do with the
fact that in quantum mechanics, unlike classical mechanics, physical properties
and variables are represented by \emph{noncommuting operators}. To be sure,
doing the $A$ and $B$ measurements of photon polarizations at different
locations, as in the usual Bell tests, guarantees that the $A_j$ operators
commute with the $B_k$ operators, but this same requirement has simply been
built into the protocol of the neon experiment.

\xb
\outl{No reason to actually carry out experiment}
\xa

Performing such an experiment would be difficult and expensive, and there is no
reason to attempt it, since by now there is vast amount of experimental
evidence that demonstrates, with high precision, the correctness of quantum
mechanics. This includes the widely publicized experiments on correlated photon
pairs, e.g. \cite{Hnao15ea,Shao15ea,Gsao15ea}, which have confirmed that
quantum theory violates Bell inequalities in the same way one would expect to
be the case were the neon experiment actually carried out.

\xb
\section{Quantum Measurements \label{sct3}}
\xa

\xb
\outl{Measurements designed to determine prior property of measured system}
\xa

\xb
\outl{Split $^{21}$Ne beam into separate beams, send these to separate 
detectors}
\xa

Quantum measurements of the sort we are interested in involve an amplification
of a microscopic quantum property in such a way to produce a macroscopic
result, the measurement \emph{outcome}, a pointer position in the archaic but
picturesque language of quantum foundations. Measurements of a beam of
$^{21}$Ne atoms could in principle be carried out in a similar way to the
famous Stern-Gerlach experiment: use magnetic (perhaps assisted with electric)
field gradients to separate the initial beam of particles into separate beams
having different properties, each identified by quantum numbers referring to
some observable. When far enough apart these beams would enter separate
detectors where individual atoms are ionized and the electron fed to an
electron multiplier resulting in a macroscopic current. Note that such a
measurement determines the property of each atom \emph{before} it is measured,
not afterwards when the measurement is over and the detector has destroyed the
atom.

\xb
\subsection{Observables and Properties \label{sbct3.1}}
\xa

\xb
\outl{Projective measurement of $F=\sum_j f_j P^j$, $\{P^j\}$ a PDI} 
\xa

This simplest sort of \emph{projective} measurement can be discussed in quantum
mechanical terms as follows. Let $F=F\ad$ be the Hermitian operator
corresponding to the physical variable (quantum observable) to be measured,
and write it in the spectral form
\begin{equation}
 F=\sum_j f_j P^j,
\label{eqn8}
\end{equation}
where the $f_j$ are eigenvalues---we assume that $f_j\neq f_k$ for $j\neq
k$---and the $P^j$ projectors onto the corresponding eigenspaces. Here the $j$
superscript of $P^j$ is a \emph{label}, not an exponent; this should cause no
confusion, because a projector is equal to its square. These projectors
satisfy the conditions
\begin{equation}
 P^j = (P^j)\ad = (P^j)^2,\quad P^jP^k = \dl_{jk}P^j,\quad I = \sum_j P^j.
\label{eqn9}
\end{equation} 
The first two equalities define a projector (orthogonal projection operator),
while the last two define the collection $\{P^j\}$ to be a \emph{projective
  decomposition of the identity $I$} (PDI). A measurement of $F$ consists in
determining which $P^j$ represents the \emph{quantum property} of the particle
being measured at a time just before the measurement takes place. The term
``property'', following von Neumann, Sec,~III.5 of \cite{vNmn32b}, corresponds
to a (closed) subspace of the Hilbert space, or its corresponding projector,
and thus refers to something which can, at least potentially, be true or false.
One should distinguish $F$, an observable or physical variable, from the
property that $F$ takes on a particular value or a range of values.%
\footnote{In this usage ``the energy of a harmonic oscillator is no greater
  than $(3/2)\hbar\om$'' is a property corresponding to a projector on a
  two-dimensional subspace, whereas ``energy'' by itself is a physical
  variable, not a property.} %
Thus a projector is the quantum counterpart of a set of points in the classical
phase space, and a PDI is the quantum analog of a probabilistic \emph{sample
  space}: a collection of mutually exclusive properties, one and only one of
which can occur in any given run of an experiment. (For more details about
measurement processes and their quantum description, see \cite{Grff17b} and
Chs.~17 and 18 of \cite{Grff02c}.)

\xb
\outl{Measuring two (in)compatible observables $F$ and $G$}
\xa

\xb
\outl{$PQ \neq QP$ $\Ra$ property ``$P$ \AND\ $Q$'' does not exist}
\xa

Suppose another observable $G=G\ad$ has the spectral form
\begin{equation}
 G = \sum_k g_k Q^k,
\label{eqn10}
\end{equation}
where the $g_k$ are its eigenvalues, and the properties $\{Q^k\}$ form a PDI.
If $F$ and $G$ commute, $FG=GF$, then every $Q^k$ commutes with every $P^j$ and
it is possible to measure $F$ and $G$ at the same time, using the PDI which is
the common refinement of $\{P^j\}$ and $\{Q^k\}$, the collection of nonzero
products $P^j Q^k$. However, if $F$ and $G$ are \emph{incompatible}, $FG\neq
GF$, then there will be some $j$ and $k$ such that $P^j Q^k\neq Q^k P^j$, and
there is no common refinement of the two PDIs, so these observables cannot be
measured in a single experiment; they must be determined in separate
experimental runs. Note that if $P^jQ^k=Q^kP^j$ the product is itself a
projector that represents the property ``$P^j$ \AND\ $Q^k$'', whereas if $P^j
Q^k\neq Q^k P^j$ neither product is a projector, so the property $P^j$ \AND\
$Q^k$ is not defined.\footnote{%
  The quantum logic of Birkhoff and von Neumann \cite{BrvN36} does assign a
  property ``$P^j$ \AND\ $Q^k$'' when the projectors do not commute, but no one
  has yet turned their quantum logic into a useful tool for reasoning in
  physical terms about microscopic quantum properties. See Sec.~4.6 of
  \cite{Grff02c} for a very simple example of one of the difficulties one runs
  into.} %
Textbooks tell us that two incompatible observables or properties cannot be
measured simultaneously, and for this there is a simple explanation (not always
given in textbooks): the simultaneous property is not represented by a
projector, and thus does not exist. Even skilled experimenters cannot measure
what is not there.

\xb
\outl{$^{21}$Ne: 4 $M_{jk}$ operators don't commute, need separate measurements}
\xa

\xb
\outl{ Calculated $\avg{S} =$ measured $M_{00} +\cdots$ a good test of SQM}
\xa

In the case of $^{21}$Ne, none of the four operators in \eqref{eqn6} commutes
with any of the other three, which means that determining their averages
requires four separate experiments. For example, $M_{01}$ has eigenvalues $+1$
and $-1$, both doubly degenerate, so the PDI contains two projectors. To find
the average $\avg{M_{01}}=\mte{\psi}{M_{01}}$ the apparatus needs to separate
the particles into two beams corresponding to these two eigenvalues, and after
a large number of runs the experimental average will be $(N_+ - N_-)/(N_+ +
N_-)$ if $N_+$ particles arrive in the $+1$ beam and $N_-$ in the $-1$ beam. A
\emph{separate experiment}, which is to say a different arrangement for
separating the incoming beams into separate beams, must be carried out for
\emph{each} of the $M_{jk}$ in order to measure its average. And since $S$ in
\eqref{eqn7} does not commute with any of the $M_{jk}$, an experimental check
of this equality in the sense of equating the average $\avg{S}$ of S, as
computed using quantum principles, with the sum of the experimental averages of
the quantities on the right side would be a rather stringent test of the
correctness of standard Hilbert space quantum mechanics.

\xb
\subsection{Quantum Measurement Model \label{sbct3.2}}
\xa

\xb
\outl{Projective measurement of $F=\sum_j f_j P^j$, $P^j = [\phi^j]$}
\xa

\xb
\outl{Initial state $\ket{\Psi_0}=\ket{\psi_0}\ot \ket{\Phi_0} = 
\sum_j c_j \ket{\phi^j}\ot\ket{\Phi_0} \in \HC_s\ot\HC_m$}
\xa

What follows is a simple quantum mechanical model of a projective measurement
of an observable $F$, \eqref{eqn8}. Additional details will be found in Chs.~17
and 18 of \cite{Grff02c}, and in \cite{Grff15,Grff17b}. In what follows we
assume that $F$ refers to a system, hereafter referred to as a `particle', with
Hilbert space $\HC_s$, while a much large Hilbert space $\HC_m$ represents the
measuring apparatus; together they constitute a closed system with Hilbert
space $\HC = \HC_s\ot \HC_m$. At an initial time $t_0$ the particle is in a
superposition of eigenstates of $F$---for simplicity assume the eigenvalues are
nondegenerate---
\begin{equation}
 \ket{\psi_0} = \sum_j c_j \ket{\phi^j},\quad P^j = [\phi^j] = \dya{\phi^j},
\label{eqn11}
\end{equation}
and the apparatus is in the `ready-for-measurement' state
$\ket{\Phi_0}$, so that the combined system is in the state
\begin{equation}
 \ket{\Psi_0} = \ket{\psi_0}\ot \ket{\Phi_0} = 
\sum_j c_j \ket{\phi^j}\ot\ket{\Phi_0}.
\label{eqn12}
\end{equation}

\xb
\outl{Times $t_0 \approx t_1 < t_2$; during $[t_1,t_2]$ unitary transformation
  $T$ }
\xa

\xb
\outl{$T (\ket{\phi^j} \ot \ket{\Phi_0}) = \ket{\Psi_2^j}$;
  $\{M^k\}$ = pointer PDI; $M^k \ket{\Psi_2^j} = \dl_{jk} \ket{\Psi_2^j}$}
\xa

Let $t_1$ be a time slightly later than $t_0$ during which there is negligible
change under unitary time evolution, so at $t_1$ $\ket{\Psi_1}$ is the same as
$\ket{\Psi_0}$. Next assume that during the time interval from $t_1$ to $t_2$
the particle and apparatus interact with each other in such a way that a
measurement process takes place, so that by $t_2$ the macroscopic quantity
representing the measurement outcome, the `pointer position', has reached its
final value.  Let $T$ be the unitary
time development operator from $t_1$ to $t_2$ ($T=\exp[-i(t_2-t_1)H]$ in the
case of a time-independent Hamiltonian $H$), and let
\begin{equation}
 \ket{\Psi_2^j} = T (\ket{\phi^j} \ot \ket{\Phi_0}),\quad
 \ket{\Psi_2} = \sum_j c_j \ket{\Psi_2^j} = T\ket{\Psi_1}.
\label{eqn13}
\end{equation}
Next assume there is a  PDI $\{M^k\}$ on $\HC$, whose
significance is that the property or projector $M^k$ corresponds to the
pointer (or whatever macroscopic variable indicates the measurement outcome)
being in position $k$, and that 
\begin{equation}
 M^k \ket{\Psi_2^j} = \dl_{jk} \ket{\Psi_2^j}. 
\label{eqn14}
\end{equation}
Thus if the particle is initially in the state $\ket{\psi_j}$ at $t_1$, its
interaction with the apparatus will result in the pointer being in position
$j$, i.e., possessing the property $M^j$, at time $t_2$, as one might expect in
the case of a projective measurement. (Note that each $M^k$, since it
represents a macroscopic quantum property, will project onto a subspace of very
high dimension, compared to which 10 raised to the power $10^{10}$ is a
relatively small number.)

\xb
\outl{Family of 3-time histories $Y^{jk} = E_0 \od E_1^j \od E_2^k$ =Qm sample
  space}
\xa

To discuss the time dependence of the measuring process when the initial
$\ket{\psi_0}$ is in a superposition---at least two of the $c_j$ in
\eqref{eqn11} are nonzero---requires the use of \emph{quantum histories},
sequences of quantum properties, represented by projectors, at successive
times.%
\footnote{See Sec.~III of \cite{Grff17b} for more details, and Chs.~8 through
  11 of \cite{Grff02c} for an extended discussion of histories and their
  probabilities.} %
For our purposes it suffices to consider histories of the form
\begin{equation}
 Y^{jk} = E_0 \od E_1^j \od E_2^k,
\label{eqn15}
\end{equation}
interpreted as meaning that the system at time $t_0$ has the property $E_0$, at
time $t_1$ the property $E_1^j$, and at time $t_2$ the property $E_2^k$. (The
symbol $\od$ denotes a tensor product, a form of $\ot$ used to separate
properties at successive times. There is no assumption that events at
successive times are related by a unitary time transformation.) Here $j$ and
$k$ are labels, and the $\{E_1^j\}$ and $\{E_2^k\}$ are PDIs. The collection
$\{Y^{jk}\}$ constitutes a \emph{family of histories} or \emph{framework}. Each
history begins with the same property $E_0$ at time $t_0$, and different
histories correspond to different events at later times. A family of histories
constitutes a quantum sample space (analogous to a collection of random walks
in classical physics) to which probabilities can be assigned using an extension
of the Born rule, provided certain consistency conditions are satisfied. In our
case the initial property is
\begin{equation}
 E_0= [\Psi_0] = [\psi_0]\ot [\Phi_0],
\label{eqn16}
\end{equation}
and we shall consider three different families or frameworks based on different
choices for the PDIs at $t_1$ and $t_2$.

\xb
\outl{Unitary framework $\FC_u: Y=[\Psi_0] \od [\Psi_0] \od [\Psi_2]$ $\ra$
superposition incompatible with pointer positions = infamous measurement
problem}
\xa

The \emph{unitary framework} $\FC_u$ contains but a single history
\begin{equation}
 \FC_u: Y=[\Psi_0] \od [\Psi_0] \od [\Psi_2],
\label{eqn17}
\end{equation}
with the projectors at $t_1$ and $t_2$ corresponding to a unitary time
development of the initial state. (Strictly speaking we should introduce a PDI
$\{[\Psi_0], I-[\Psi_0]\}$, $I$ the identity operator, at $t_1$, but the
extended Born rule assigns zero probability to the second of these
possibilities, so it can be ignored; similarly a PDI $\{[\Psi_2], I-[\Psi_2]\}$
at time $t_2$.) The trouble with the family $\FC_u$ is that when two or more of
the $c_j$ are nonzero, the state $\ket{\Psi_2}$ is a coherent superposition of
states that correspond to different pointer positions, and hence the
corresponding property $[\Psi_2]$ does not commute with projectors representing
different positions of the pointer; the two are incompatible, and trying to
combine them will give a meaningless result, as noted earlier in the case of
incompatible observable $F$ and $G$. We have arrived at the infamous
measurement problem of quantum foundations, or, in popular parlance,
Schr\"odinger's cat.

\xb
\outl{Family $FC_1: Y^k =[\Psi_0] \od [\psi_0]\ot[\Phi_0] \od M^k$; $Y^k$ has
  Born rule probability}
\xa

This difficulty can be avoided by using in place of $\FC_u$ a family
\begin{equation}
 \FC_1: Y^k =[\Psi_0] \od [\psi_0] \od M^k,
\label{eqn18}
\end{equation}
where by physicists' convention $[\psi_0]$ at $t_1$ stands for $[\psi_0]\ot
I_m$ on the full Hilbert space, and histories with $I-[\psi_0]$ at $t_1$ have
been omitted since they have zero probability. The use of $[\psi_0]$ rather
than $[\Psi_0]$ as in \eqref{eqn17} serves to focus attention on the particle
at time $t_1$. The $k$'th history $Y^k$ ends in the pointer position $M^k$ at
time $t_2$, and the extended Born rule assigns to this outcome a probability
\begin{equation}
 \Pr(Y^k) = \mte{\Psi_2}{M^k} = |c_k|^2 = \mte{\psi_0}{P^k}.
=|\inpd{\phi^k}{\psi_0}|^2.
\label{eqn19}
\end{equation}
The final expression on the right is the formula students learn in an
introductory course.

\xb
\outl{Open black box: $\FC_2: Y^{jk}=[\Psi_0] \od P^j \od M^k$ $\ra$
$\Pr(Y^{jk}) = \dl_{jk} |c_k|^2$\\ $\Pr([\phi^j]_1 \vbl [M^k]_2) = \dl_{jk}$}
\xa

One can go a step further in opening the black box by using the framework
\begin{equation}
  \FC_2: Y^{jk}=[\Psi_0] \od [\phi^j] \od M^k,
\label{eqn20}
\end{equation}
where $[\phi^j]$ (i.e., $[\phi^j] \ot I_m$) at $t_1$ means the particle has the
property $[\phi^j]$, while nothing is said
about the state of the apparatus. It is easily shown that the consistency
conditions for this family are satisfied, and the extended Born's rule assigns
probabilities
\begin{equation}
 \Pr(Y^{jk}) = \dl_{jk} |c_k|^2=|\inpd{\phi^k}{\psi_0}|^2.
\label{eqn21}
\end{equation}
This agrees with \eqref{eqn19}, but provides additional information, namely the
conditional probabilities (where subscripts $1$ and $2$ identify the time):
\begin{equation}
 \Pr([\phi^j]_1 \vbl [M^k]_2) = \dl_{jk} = \Pr( [M^j]_2  \vbl [\phi^k]_1),
\label{eqn22}
\end{equation}
assuming $c_k\neq 0$. The first says that if the measurement outcome (pointer
position) is $k$ at $t_2$, then at the earlier time $t_1$, \emph{before} the
measurement took place, the particle had the corresponding microscopic property
$[\phi^k]$. In other words, a projective measurement of this sort reveals a
prior property of the measured system when one uses an appropriate quantum
description that allows for this possibility. Herein lies the key difference
between $\FC_2$, in which the different $[\phi^k]$ make sense at $t_1$, and
$\FC_1$, where they do not, since $[\psi_0]$, assuming at least two of the
$c_j$ in \eqref{eqn13} are nonzero, does not commute with the relevant
$[\phi^k]$.

\xb
\outl{$\ket{\psi_0}$ used at time $t_1$ in $\FC_2$ is \emph{pre-probability},
  not property}
\xa

In addition, since in $\FC_2$ $[\phi^k]$ occurs at an \emph{earlier time} than
the measurement outcome $M^k$, the second equality in \eqref{eqn22} allows one
to identify the earlier $[\phi_k]$ as the \emph{cause} of the later $M^k$.
This is the way an experimenter will normally think about the operation of a
measurement apparatus; e.g., it is the arrival of a photon which caused the
photodetector to produce a click, not vice versa.
Note that the superposition state $\ket{\psi_0}$, whereas it does not appear at
time $t_1$ in $\FC_2$, can nonetheless be used, as in \eqref{eqn21}, for
calculating the probabilities assigned to the different properties $[\phi^k]$
at this time. A wavefunction or ket used in this manner is referred to as a
`pre-probability' in Sec.~9.4 of \cite{Grff02c}. This role as a calculational
tool should be carefully distinguished from its use as a quantum property, as
in \eqref{eqn18}.

\xb
\outl{Apparatus calibration enables retrodiction: outcome $\Ra$ earlier Qm
  property }
\xa

A careful experimenter will want to check that the measurement apparatus built
to measure a particular observable is functioning properly. One check is
calibration: if the device has been built to measure $F$ in \eqref{eqn8}, then
for each $j$ send in a stream of particles known to have the property $P^j$ and
check that the pointer always ends up at position $j$. Once the device has been
calibrated, the experimenter will normally assume that if a particle whose
property is unknown arrives at the detector and the pointer points at $j$, then
the particle earlier had the property $P^j$. Thus the earlier property can be
inferred or retrodicted from the measurement outcome.

\xb
\outl{Retrodiction valid even in particle initially in superposition state}
\xa

But what if the particle was initially prepared in a superposition
$\ket{\psi_0}$ of states corresponding to different values of $j$? The use of
the framework $\FC_2$ shows that such an inference remains valid. If the same
initial state is used in a successive runs of the experiment, the outcomes will
be different, with probabilities given by the usual formula \eqref{eqn19}. It
is not meaningful to ask, ``Did the particle have the property $[\psi_0]$ or
the property $[\phi^k]$ prior to the measurement?'', because the projectors do
not commute. But if the question is: ``Which among
the $[\phi^j]$ was the property possessed by the particle just before it
reached the apparatus,'' then the answer is given by using the framework
$\FC_2$ leading to the formula \eqref{eqn22}. Inferences of this sort are made
all the time by experimenters, and it is to be regretted that this ``common
sense'' understanding of quantum measuring processes is not explained in
introductory textbooks.

\xb
\outl{No one of $\FC_1. \RC_2, FC_3$ is the \emph{right} framework; choice
  $\lra$  question being asked}
\xa

We have employed three distinct frameworks or families of histories, $\FC_u$,
$\FC_1$, and $\FC_2$ in order to describe what goes on in a projective
measurement. Which is the \emph{right} framework? That depends on the question
one wishes to address. If one is interested in relating the measurement outcome
to the quantity it was designed to measure, $\FC_2$ is the right framework,
because it contains the corresponding microscopic events. These events are not
simply absent from $\FC_u$ and $\FC_1$; in those families they have no meaning,
because the $[\phi^j]$ are incompatible with the projectors used in $\FC_u$ and
$\FC_1$ at time $t_1$. On the other hand, were one interested in whether the
particle was perturbed on its way from an initial preparation to the time $t_1$
just before the measurement took place, a PDI at $t_1$ that included the state
that evolved unitarily from the initial preparation would be appropriate. It is
always a mistake to try and answer a question about a quantum property using a
framework in which it is meaningless.

\xb
\outl{Different frameworks answer different questions}
\xa

Different incompatible frameworks are used in quantum mechanics for answering
different questions, and it is important to note that when a particular setup
allows for several alternative incompatible frameworks, the answer provided by
one of them to a question properly posed (in quantum terms) is not invalidated
by the existence of alternative frameworks. Instead, there is a general
consistency argument, see Ch.~16 of \cite{Grff02c}, that using alternative
frameworks will never lead to contradictory results, i.e., some property $P$ is
true (probability 1) in one framework and false (probability 0) in another
framework. Numerous quantum paradoxes represent apparent violations of this,
but when examined they always involve some combination of arguments carried out
by combining results from incompatible frameworks. Thus a central principle of
CH is the \emph{single framework rule}: valid quantum reasoning requires that
different parts of an argument can all be embedded in, or expressed using, a
single overall framework. The choice of which framework to use will depend upon
which questions one wishes to answer. If one wants to assign probabilities to
measurement outcomes it is necessary to employ a quantum description or
framework in which the different macroscopic outcomes make sense: thus $\FC_1$
or $\FC_2$, rather than $\FC_u$, for the example discussed above. If one wants
to relate the measurement outcome to the corresponding prior microscopic
property that was measured, the framework must be one in which those properties
make sense, $\FC_2$ rather than $\FC_u$ or $\FC_1$.

\xb
\subsection{Quantum Particle In Different Locations? \label{sbct3.3}}
\xa

\xb
\outl{Projector $\hat R$ for particle in region $R$}
\xa

Can a quantum particle be in two different locations at the same time? To
address this we first need to say what it means for a quantum particle to have
the property that it is in some region of space $R$. That property is
represented by a projector $\hat R$ whose action on the position-space
wavefunction $\psi(\rB)$ is given by
\begin{equation}
 \hat R\psi(\rB) = \begin{cases}
 \psi(\rB) \text{ if $\rB \in R$}\\
  0 \text{ otherwise.}
\end{cases}
\label{eqn23}
\end{equation}
That is, it sets $\psi(\rB)$ to zero when $\rB$ is not in $R$, but otherwise
leaves it unchanged. The projector for the particle to be simultaneously in two
regions $R_1$ and $R_2$ is $\hat R_1 \hat R_2 = \hat R_2 \hat R_1$. If
the regions $R_1$ and $R_2$ do not overlap, this product is zero, which means
the corresponding property cannot occur. Thus if `two places' is understood as
two regions in space that do not overlap, the particle cannot be in both of
them at the same time.

\xb
\outl{Particle location measurement $\ra$only one place $\Ra$
  particle not in 2 places}
\xa

\xb
\outl{$\psi(\rB)$ can be considered a pre-probability}
\xa

Once one understands that projective quantum measurements can be understood as
measuring prior properties, the same conclusion follows from the textbook
statement that even if a particle has a spread-out wavefunction, a measurement
of position will find it in only one place. Thus if the support of the particle
wavefunction is in the union $R = R_1 \cup R_2$
\ca
\begin{equation}
 R = R_1 \cup R_2
\notag
\end{equation}
\cb%
of two nonoverlapping regions $R_1$ and $R_2$, a
position measurement will reveal its presence in one but not in the other, and
its position just prior to measurement will be in the region indicated by the
measurement outcome.
Note that the \emph{property} $[\psi]$, the Hilbert space projector that
corresponds to the wavefunction $\psi(\rB)$, will not commute with either of
projectors $\hat R_1$ or $\hat R_2$ associated with these two regions. assuming
the support of $\psi(\rB)$ is not confined to one or the other. Thus in
calculating the probabilities that the particle will be in (thus measured to be
in) $R_1$ or $R_2$ one must understand $\psi(\rB)$ to be a
\emph{pre-probability}; assuming it is normalized, $\rho(\rB) = |\psi(\rB)|^2$
is a probability density which can be integrated over $R_1$ or $R_2$ to find
the probability that the particle is in one of these regions, or that an
appropriate measurement will find it there. Thus implicit in our discussion is
a framework analogous to $\FC_2$ in \eqref{eqn20}.

\xb
\outl{Application: double-slit experiment, regions $R_j$, $j=1,2$ include slit
  $j$}
\xa

As a particular application one can think of the case of a double-slit
experiment, and let $R_1$ and $R_2$ be nonoverlapping regions, where $R_1$
includes the first slit and its vicinity, but not the second slit, while $R_2$
is the vicinity of the second slit, but excludes the first. Suppose that at a
particular time the wave representing the particle is in the union of $R_1$ and
$R_2$. If detectors are placed immediately behind each slit, detection will
show that the particle was in one of these regions, not both. If, on the other
hand, the particle/wave emerging from the slits is undisturbed as it proceeds
towards the distant interference region where there are a large number of
detectors which detect its position at a later time, then it is correct to say
that at the earlier time the particle was in the region $R$, but introducing
the separate regions $R_1$ and $R_2$ into the quantum description at this time
will violate the consistency conditions required to assign probabilities,
making ``Which slit did it pass through?'' a meaningless question.%
\footnote{There is an alternative framework in which the particle passes
  through a definite slit, but the detectors in the later interference region
  end up in a macroscopic quantum superposition (Schr\"odinger cat) state. One
  can understand why this framework has little appeal for understanding real
  experiments!} %
For more details see Ch.~13 of \cite{Grff02c}.

\xb
\outl{Particle in $R=R_1\cup R_2$ $\not\Ra$ particle in $R_1$ or $R_2$}
\xa

\xb
\outl{Analogy: Energy of harmonic oscillator two lowest levels}
\xa

But, the reader may ask, if the particle was in $R=R_1\cup R_2$, 
does that not immediately imply that it was either in $R_1$ or else it was in
$R_2$? That would represent good classical reasoning, but it need not hold in
the quantum world. To see why it can fail, consider a different situation: a
quantum harmonic oscillator in which the possible energies are
$(n+1/2)\hbar\om$ with corresponding (orthogonal) eigenstates $\ket{n}$,
$n=0,1,\cdots$. Consider the two-dimensional subspace spanned by $\ket{0}$ and
$\ket{1}$ whose projector is $P=[0]+[1]$. If the oscillator is in either of the
two energy eigenstates $\ket{0}$ or $\ket{1}$, it possesses the property $P$.
However, a superposition state $\ket{\chi} =(\ket{0} + \ket{1})/\st$ also lies
in this 2-dimensional subspace, but does \emph{not} possess either property
$[0]$ or $[1]$, as it does not have a well-defined energy. Similarly, a quantum
particle passing through a double-slit system cannot, in general, be said to
pass through a particular slit.

\xb
\section{Classical Hidden Variables \label{sct4}}
\xa

\xb
\outl{Claim of nonlocality for any present or future theory}
\xa

There are a large number of published derivations of Bell inequalities, and it
has even been claimed 
\cite{dEsp06, Mdln11c, Nrsn17}
that \emph{any local} theory of the world, present or future, \emph{must} lead
to inequalities of this sort. That is, the experimental violations of Bell
inequalities not only imply that the quantum world is nonlocal, but any future
theory that gives results in agreement with these experiments will involve the
same nonlocality. It is therefore useful to say a few words about what is wrong
(from the perspective of Hilbert-space quantum mechanics) with the assumptions
made in typical derivations of Bell inequalities, and why the aforementioned
claim is false.

\xb
\outl{Bell derivations use \emph{factorization condition}
 $\Pr(A,B|a,b) = \sum_\lm \cdots$.}
\xa

\xb
\outl{Symbols defined.  $\lm=$ hidden variable, a common cause}
\xa

It will suffice to focus on the \emph{factorization condition}, which always
appears in some form or another in a derivation of the CHSH or other Bell
inequalities:
\begin{equation}
 \Pr(A,B\vbb a,b) = \sum_\lm \Pr(A \vbb a,\lm) \Pr(B \vbb b,\lm) \Pr(\lm).
\label{eqn24}
\end{equation}
The symbols entering this expression have the following significance. Alice and
Bob, who are far away from each other, are measuring pairs of particles
produced at a common source. The outcome (pointer position) of Alice's
measurement is $A$ given the setting $a$ of her apparatus, which determines the
type of measurement being performed. Likewise, $B$ and $b$ refer to the outcome
and setting for Bob's measurement. On the right side of \eqref{eqn24} the
``hidden variable'' $\lm$ determines, in a probabilistic sense, the dependence
of $A$ on $a$ and of $B$ on $b$. (One can replace the sum over $\lm$ with an
integral; it makes no difference.). The equation \eqref{eqn24} expresses
locality in the sense that if Alice and Bob are far from each other, the choice
of $a$ and the resulting outcome $A$ should not influence $B$, nor the choice
of $b$ influence $A$, as long as $\lm$, a ``common cause'', is held fixed.

\xb
\outl{Apply factorization to single term $M_{00} =A_0B_0$ in $S=\cdots$}
\xa

To better understand the connection of such hidden variables with Hilbert space
quantum mechanics, consider applying \eqref{eqn24} to just one of the terms on
the right side of \eqref{eqn7}, say $M_{00}=A_0 B_0$, whose average in the
state $\ket{\psi}$ we wish to evaluate using a proper quantum-mechanical
calculation. Let us set $a=0,\, b=0$, and since they are fixed, 
drop them from both sides of \eqref{eqn24}, which becomes
\begin{equation}
 \Pr(A_0,B_0) = \sum_\lm \Pr(A_0 \vbb \lm) \Pr(B_0 \vbb \lm) \Pr(\lm).
\label{eqn25}
\end{equation}
Here $A_0$ and $B_0$ are defined in \eqref{eqn5}, with eigenvalues $\pm1$, so
can be written in the form, see \eqref{eqn8}:
\begin{equation}
 A_0 = P_+ - P_-,\quad B_0 = Q_+ - Q_-,
\label{eqn26}
\end{equation}
using the two commuting PDIs $\{P_+,P_-\}$ and $\{Q_+,Q_-\}$. One can 
think of the arguments of $\Pr(A_0,B_0)$ as the eigenvalues of these operators,
and thus  \eqref{eqn25} as the set of four equations, one for each $p$ and $q$,
\begin{equation}
 \Pr(P_p Q_q) = \sum_\lm \Pr(P_p \vbb \lm) \Pr(Q_q \vbb \lm) \Pr(\lm), 
\label{eqn27}
\end{equation}
which assigns probabilities to the projectors $P_p Q_q$ that together
constitute the quantum sample space that is the common refinement 
of the PDIs used in \eqref{eqn26}. 
Identifying \eqref{eqn24} with \eqref{eqn25} is not completely trivial, since
in the former $A$ and $B$ represent scalar quantities, measurement outcomes of
$+1$ and $-1$, whereas in \eqref{eqn25} $A_0$ and $B_0$ refer to the
eigenvalues $+1$ and $-1$ of quantum operators, and \eqref{eqn27} to the
corresponding eigenspaces. This identification is correct provided projective
measurements reveal pre-existing values, as explained in Sec.~\ref{sct3}.

\xb
\outl{Identify HV $\lm$}
\xa

The two sides of \eqref{eqn27} will be equal if we let the hidden variable
$\lm$ take on one of the four values $++$, $+-$, $-+$, $--$, given by the pair
$pq$, and use conditional probabilities
\begin{equation}
  \Pr(P_p \vbb p'q) = \dl_{pp'},\quad \Pr(Q_q \vbb pq') = \dl_{qq'}
\label{eqn28}
\end{equation}
together with 
\begin{equation}
 \Pr(\lm=pq) = \mte{\psi}{P_p Q_q}.
\label{eqn29}
\end{equation}
Inserting these in the right side of \eqref{eqn27} makes it equal to
$\mte{\psi}{P_p Q_q}$, the Born rule for $\Pr(P_p Q_q)$. Thus we have
a particular quantum application of \eqref{eqn24} in the case
$a=0,\,b=0$.

\xb
\outl{Need same HV for all $j,k$ as BI derivation assumes $\lm$ independent of
  $a,b$}
\xa

What works for $a=0,\,b=0$, the $M_{00}$ term in \eqref{eqn7}, will work
equally well for any of the other terms; one simply has to use appropriate
choices for the PDIs $\{P_p\}$ and $\{Q_q\}$. But we know that the quantum
average of the quantum $S$ in \eqref{eqn7} can exceed the classical CHSH bound
in \eqref{eqn3}. Why is this? The trouble arises because if, for example, we
consider $M_{10}$ in place of $M_{00}$ we will need a different choice for
$P_+$ and $P_-$ when $A_0$ in \eqref{eqn26} is replaced with $A_1$, with which
it does not commute, and this means changing the definition, or at least the
physical meaning, of $\lm$. But derivations of Bell inequalities always assume
that $\lm$, which is supposed to represent an earlier state of the measured
particles, does \emph{not} depend on the choices of $a$ and $b$ made by Alice
and Bob, so changing it is not allowed.

\xb
\outl{Bell inequality derivations do not accommodate noncommutation}
\xa

Might there be some different choice for $\lm$ that evades this difficulty? Not
likely, given the large number of careful analyses that show that
\eqref{eqn24}, with $\lm$ independent of $a$ and $b$, leads inexorably to the
CHSH inequality, which is \emph{not} satisfied by the correct quantum average
of $S$ in \eqref{eqn7}. What the foregoing analysis suggests is that the
fundamental problem with such derivations is that they do not take proper
account of the possible \emph{noncommutativity} of quantum projectors
representing the quantum properties of interest. And since this failure applies
to $^{21}$Ne, locality cannot be an issue.

\xb 
\outl{Basic problem: factorization assumes $\lm$ is element of Cl
  sample space} 
\xa

To summarize, the fundamental difficulty with the factorization condition
\eqref{eqn24} is that it assumes a \emph{single} sample space of
mutually-exclusive possibilities, independent of $a$ and $b$, with elements
labeled by $\lm$. This would be quite appropriate for a classical system where
there is a single phase space and the sample space employs non-overlapping
subsets of this phase space. But a quantum Hilbert space allows incompatible
samples spaces, different PDIs with projectors that do not commute, and
therefore lack a common refinement. Thus the usual derivations of CHSH and
other Bell inequalities employ \emph{classical} physics to discuss
\emph{quantum} systems, so it is not surprising when these inequalities fail to
agree with quantum predictions, or the experiments that confirm these
predictions.

\xb
\section{Einstein Podolsky Rosen (EPR)\label{sct5}}
\xa

\xb
\subsection{Bohm Version of EPR \label{sbct5.1}}
\xa

\xb
\outl{Singlet state, correlated measurement outcomes}
\xa

While the mistake associated with the claim that the violation of Bell
inequalities implies nonlocality in the quantum world should be evident from
the neon example of Sec.~\ref{sbct2.2}, and from the use of classical hidden
variables for deriving these inequalities, Sec.~\ref{sct4}, there are
useful lessons to be learned from considering the original
Einstein-Podolsky-Rosen (EPR) argument \cite{EnPR35}, where locality was simply
assumed, using the simplified version introduced by Bohm, Ch.~22
of \cite{Bhm51}. Two spin-half particles, $a$ and $b$, are prepared in the
spin-singlet state
\begin{equation}
 \ket{\psi_s} = \left( \ket{0}_a\ot\ket{1}_b -
\ket{1}_a\ot\ket{0}_b \right)/\st,
\label{eqn30}
\end{equation}
with $\ket{0}$ and $\ket{1}$ the $+1/2$ and $-1/2$ (in units of $\hbar$)
eigenstates of $S_z$. Particle 1 is sent to Alice and 2 to Bob, who can then
carry out measurements of the same or different components of spin angular
momentum. If they measure the same component, say $S_w$, where $w$ could be $x$
or $z$ or any other direction in space, the results will be
opposite: if Alice observes $+1/2$ Bob will find $-1/2$, or $+1/2$ if
Alice observes $-1/2$.

\xb
\outl{No Hilbert subspace corresponds to ``$S_x=+1/2$ \AND\ $S_z = -1/2$''}
\xa

Recall that the Hilbert space of a spin-half particle is two-dimensional, and
thus the PDI associated with any spin component $S_w$ consists of two
projectors onto pure states. Neither projector associated with $S_x$
commutes with either of the projectors associated with $S_z$, and
consequently there is no subspace of the Hilbert space which can represent
simultaneous values of both $S_x$ and $S_z$. Hence expressions like
``$S_x=+1/2$ \AND\ $S_z = -1/2$'' are meaningless,\footnote{%
  In quantum logic such a conjunction is a property that is always
  false (the zero-dimensional subspace).} %
and the same holds for any two distinct components of angular momentum.

\xb
\subsection{The Counterfactual Argument \label{sbct5.2}}
\xa

\xb
\outl{Apparatus measures either $S_z$ or $S_x$; choice made just before particle
  arrives }
\xa

While, for reasons given above, $S_x$ and $S_z$
for a spin-half particle cannot be measured simultaneously, it is possible in
principle to design an apparatus to measure \emph{either} $S_x$ \emph{or} $S_z$,
with the choice between the two made just before the particle enters the
measuring device. (E.g., a small region with a uniform magnetic field in the $y$
direction placed just in front of the apparatus can cause $S_x=\pm 1/2$ to
precess into $S_z=\pm1/2$, turning an $S_z$ into an $S_x$ measurement; this
field can be switched on or off just before the arrival of the particle.)

\xb
\outl{Ctfl: $S_z$ measured, $S_x$ could have been measured: $\Ra$ QM
  incomplete?}
\xa

\xb
\outl{References for additional discussions of counterfactuals, locality}
\xa

Suppose that with the $S_z$ setting Alice finds $S_z=+1/2$ during a particular
run. One can imagine that Alice \emph{could} have chosen the $S_x$ setting, and
in that case \emph{would have} obtained either $S_x=+1/2$ or $-1/2$, we do not
know which. Does it not follow that the particle had \emph{both} a definite
$S_z$ value revealed by the later measurement\emph{and} a specific $S_x$
component, the one that Alice \emph{would} have learned \emph{had} she measured
$S_x$ rather than $S_z$, a choice which she \emph{could have made} at the very
last instant before the particle reached the apparatus? The italicized words
indicate that this is a \emph{counterfactual} argument: it combines what
actually happened with what \emph{would have} happened in a similar but
different situation. (For a discussion of consistent ways to discuss
counterfactuals within quantum mechanics, see \cite{Grff99} or Ch.~19 of
\cite{Grff02c}, and for an application to (non)locality issues, the interchange
in \cite{Stpp12,Grff12b}.) Doesn't this prove that the quantum Hilbert space
provides but an \emph{incomplete} description of physical reality? The reader
familiar with their original paper will notice the similarity with EPR's
argument, which also contains the (implicit) assumption that if one measured
one observable, one could very well have measured a different, incompatible
observable.

\xb
\outl{Reverse EPR: QM complete $\Ra$ separate runs for incompatible observables;
locality has nothing to do with the matter}
\xa

However, one can just as well run the EPR argument in reverse. Given a
classical situation where all observables (by definition) commute, or a quantum
situation with two \emph{commuting} observables, $FG=GF$, it makes perfectly
good sense to ask: Suppose $F$, \eqref{eqn8} was measured with the result
indicating, say, the property $P_2$, what \emph{would have happened} in this
instance \emph{if instead} $G$, \eqref{eqn10} had been measured, i.e., what is
the probability that the measurement would have revealed a property $Q_k$?
Given some initial state the joint probability distribution corresponding to
the common refinement, the PDI composed of the nonzero $P_jQ_k$, can be
computed, and from it a conditional probability $\Pr(Q_k\vbl P_2)$, which is
then a sensible (in general, probabilistic) answer to the counterfactual
question. But when the projectors do not commute this cannot be done, and then,
as noted earlier, $F$ and $G$ must be measured in separate experiments, and
there is no reason to suppose that the value of $F$ revealed in one experiment
has anything to do with the value of $G$ obtained in a different, independent
experiment. In other words, the \emph{completeness} of Hilbert space quantum
mechanics, which makes \emph{impossible} the simultaneous measurement of $S_x$
and $S_z$, as there is nothing in the Hilbert space that corresponds to a joint
property, undermines the counterfactual assumption that when Alice measured
$S_z$ she \emph{could have} measured $S_x$ \emph{in the same run}. Were $S_x$
to have been measured, it would have to have been in a different run, and there
is no reason why the value of $S_z$ measured in one run will somehow be related
to the value of $S_x$ measured in a different run.

\xb
\outl{Ctfl argument implicit in EPR is blocked if QM is complete}
\xa

Hence the counterfactual notion which enters, at least implicitly, the EPR
argument is blocked as soon as one assumes, contrary to EPR, that Hilbert space
quantum theory is complete, and there are no additional hidden variables. Given
that attempts to supplement the quantum Hilbert space with hidden variables
have thus far failed---as shown most clearly by experiments confirming the
(Hilbert space) quantum violations of Bell inequalities
\cite{AsGR81,Hnao15ea,Shao15ea,Gsao15ea}, it would seem that the original EPR
argument, that (Hilbert space) quantum mechanics is incomplete, fails.
Locality, or its absence, has nothing to do with the matter: the issue is what
measurements carried out on a \emph{single} particle in a \emph{single}
location can tell one about the properties of \emph{that} particle.

\xb
\subsection{Quantum Common Cause \label{sbct5.3}}
\xa

\xb
\outl{Nonlocality belief arises from correlations without common cause}
\xa

\xb
\outl{Will show there is a \emph{quantum} common cause}
\xa

\xb
\outl{Photon experiments use common source/cause to define coincidences}
\xa

\xb
\outl{Need argument for common source of polarizations}
\xa

As noted in Sec.~\ref{sct1}, one reason for the belief in instantaneous
nonlocal quantum influences is that quantum theory predicts, and experiment
confirms, the existence of \emph{correlations} which violate Bell inequalities,
and thus cannot be explained by a common cause based on classical hidden
variables. However, opening the black box and applying consistent quantum
principles provides an explanation for the correlations in terms of local
\emph{quantum} common causes.
Experiments that test Bell inequalities using entangled photon pairs already
assume a common cause in the sense that pairs of photons produced at the source
in the same, rather than a different, down conversion event are identified
using their arrival times. All that is needed in addition is an argument that
the polarizations measured later were also created in the same (local) event.

\xb 
\outl{A,B (spin) measurements $\lra$ prior properties traced back to common
preparation $\Ra$ correlation via common cause, not evident in `collapse'
framework} 
\xa

Here we employ the principle discussed in Sec.~\ref{sct3}, that measurements of
a suitable sort can be interpreted, by using a suitable framework, as revealing
prior properties of the measured system. Reverting to spin-half language, if
Alice's apparatus is set to measure $S_z$ for particle $a$ and the outcome
corresponds to, say, $S_z=-1/2$, she can conclude that particle $a$ possessed
this property just before the measurement took place, and, assuming it was not
perturbed on its way to her apparatus, at all previous times following the
initial preparation. The same applies to Bob's measurement of $S_z$ for
particle $b$. Thus by applying the Born rule right after the two particles are
prepared in the singlet state \eqref{eqn30}, one sees that the probability that
particles 1 and 2 have the same z component of spin is zero, and the two
possibilities for opposite $S_z$ values each has a probability of 1/2. A similar
argument using an appropriate framework applies to the case where Bob
measures $S_w$ for an arbitrary direction $w$. The probabilities for the
correlations predicted by using what might be called a `measurement' framework,
in which both measurement outcomes are traced all the way back to the source,
are exactly the same as those predicted by textbook quantum theory using
wavefunction collapse in a `collapse' framework, Sec.~\ref{sct6},
in which the entangled singlet state persists right up to the instant
before one of the measurements. There is no reason that the Born rule can only
be applied when a measurement takes place; this mistaken notion has been one of
the reasons for the lack of progress in quantum foundations in resolving its
infamous `measurement problem'.

\xb
\outl{Collapse framework does not invalidate inferences in
prior property framework}
\xa

\xb
\outl{$M_{00}$ measurement in Sec.~\ref{sct4}, illustrates `quantum cause'}
\xa

As noted in Sec.~\ref{sbct3.2}, inferences obtained in one framework are not
invalidated by the existence of alternative frameworks. The `collapse'
framework, which treats the entangled state as a property right up until the
measurement takes place, precludes any discussion during that time period of
spin states of the individual particles---see the comments in
Sec.~\ref{sct6}---thus concealing the fact made obvious in the `measurement
framework', in which measurements reveal prior properties, that the quantum
correlations between measurement outcomes have an explanation in terms of a
(quantum) common cause.
The reader may also find it helpful to consider the discussion of the
measurement of $M_{00}$ in Sec.~\ref{sct4}, where proper use was made of a
genuinely quantum `hidden variable' $\lm$, as an example of a `quantum cause',
in the same sense as that employed here.

\xb
\outl{Alice' measurement of $a$ does not affect particle $b$, but her knowledge
of the initial $\ket{\psi_s}$ allows her to infer a property of $b$}
\xa

Alice's choice of measurement on particle $a$ has no influence at all on Bob's
particle $b$ and whatever measurements may be carried out on it. However, her
knowledge of the outcome of a measurement of a particular component of angular
momentum allows her to infer a property possessed by particle $a$ before the
measurement took place. Combined with what she knows about the preparations
protocol, in particular the initial state $\ket{\psi_s}$, this allows her to
infer something about particle $b$, from which she can also infer the
probability of the outcome of a measurement of particle $b$. Thus if particle
$a$ is measured to have $S_z=-1/2$, Alice can assign an $S_z=+1/2$ property to
particle $b$ and predict with certainty the outcome of Bob's measurement of
$S_z$, or assign a probability to the outcome if Bob instead measures some
other component of spin angular momentum.

\xb
\outl{Classical analogy: colored slips of paper sent to Alice, Bob}
\xa

The following classical analogy may help in understanding this.
Charlie inserts red and green slips of paper into two identical, opaque
envelopes; then chooses one at random and mails it to Alice in Atlanta, and the
other to Bob in Boston. From her knowledge of the preparation protocol Alice,
upon opening her envelope and seeing the color of the slip of paper it contains,
can immediately infer the color of the paper in Bob's envelope, whether or not
he has already opened it or will open it at a later time. No magic or
mysterious long-range influence is needed to understand how this works, and the
same is true of its quantum analog.

Granted, this classical analogy does not cover all possibilities present in the
quantum case; in particular the situation in which Alice measures one component
of spin angular momentum and Bob a different component. However, it is still
correct to say that from the $S_z$ outcome of her measurement and her knowledge
of the initial preparation, Alice can assign (conditional) probabilities to the
outcomes of a measurement by Bob in the $S_x$ or any other basis, and this
possibility has nothing to do with her measurement having some mysterious
effect upon Bob's particle.

\xb
\section{Wavefunction Collapse and Einstein Locality \label{sct6}} 
\xa

\xb
\outl{EPR $\Pr(a^j,b^k) = \mte{\psi_s}{\,[a^j]\ot [b^k]\,}$}
\xa

Spin-spin correlations in the Bohm version of EPR are usually calculated
by one of two closely-related methods. Let us suppose that Alice and Bob carry
our measurements in the orthonormal bases $\{\ket{a^0},\ket{a^1}\}$ and
$\{\ket{b^0},\ket{b^1}\}$, respectively. The joint probability distribution
for an initial state $\ket{\psi_s}$ can be computed using the standard formula
\begin{equation}
\Pr(a^j,b^k) = \mte{\psi_s}{\,[a^j]\ot [b^k]\,}.
\label{eqn31}
\end{equation}
The discussion in Sec.~\ref{sbct3.2} justifies thinking of $\ket{\psi_s}$
as a pre-probability, and identifying $[a^j]$ and
$[b^k]$ as properties of the $a$ and $b$ particles prior to the measurement, 
the point of view adopted in the common cause discussion in Sec.~\ref{sbct5.3}.

\xb
\outl{$\Pr(a^0,b^k)$ by using `collapse' by measurement of $a$ to give
$\Pr(b^k\vbl a^0)$ }
\xa

An alternative approach which yields the same joint probabilities employs
\emph{wavefunction collapse}. Assume that Alice's measurement is carried out
first, and the outcome corresponds to $[a^0]$. This is thought of as
``collapsing'' the wavefunction $\ket{\psi_s}$ to a new state
\begin{equation}
 \ket{\psi_c^0} = [a^0]\, \ket{\psi_s}/\sqrt{\mte{\psi_s}{\,[a^0]\,}},
\label{eqn32}
\end{equation}
(where $[a^0]$ stands for $[a^0]\ot I_b$). The (conditional)
probability that Bob's measurement outcome will correspond to $[b^k]$ is then
computed using the collapsed state:
\begin{equation}
 \Pr(b^k\vbl a^0) = \mte{\psi_c^0}{\,[b^k]\,}.
\label{eqn33}
\end{equation}
When multiplied by $\Pr(a^0) = \mte{\psi_s}{\,[a^0]\,}$ this gives the result
in \eqref{eqn31}.

\xb
\outl{The `collapse' from Alice's measurement of $a$ $\not\Ra$ effect on
  particle $b$}
\xa

\xb
\outl{Confusion arising from Schr\"odinger steering}
\xa

There is nothing wrong with this collapse procedure for obtaining the result
in \eqref{eqn31}. However, as noted earlier in Sec.~\ref{sbct5.3}, Alice's
measurement has no effect upon Bob's particle. Thus treating the collapse
process in which $\ket{\psi_s}$ is replaced by $\ket{\psi_c^0}$, as an actual
physical process in which Alice's measurement has somehow altered a property of
particle $b$, is incorrect, and this error has given rise to a great deal of
confusion, starting with EPR and extending up to more recent discussions of
\emph{steering}, e.g. \cite{WsJD07,CvSk17,Tsao18}, a term originating with
Schr\"odinger \cite{Schr36} and expressing the idea that if Alice and Bob share
an entangled state, Alice's measurement may be able to alter Bob's particle.

\xb
\outl{$\ket{\psi_s}$ = pre-probability; $[\psi_s]$ inconsistent with individual
$a$, $b$ properties}
\xa

\xb
\outl{Does Alice measurement affect $b$?---requires appropriate framework.
  Answer: no effect. Special case of Principle of Einstein Locality}
\xa

\xb
\outl{Entangled state $\not\Ra$ one system influences the other}
\xa

The mistake arises from a misunderstanding of the collapse framework. When
$\ket{\psi_s}$ is employed as a pre-probability, as in \eqref{eqn31}, it cannot
be identified with a physical property of either particle $a$ or $b$, since the
corresponding projector $[\psi_s]$ does not commute with any nontrivial
property of either particle. (The trivial properties are the identity projector
$I$, always true, and the zero projector, always false.) Therefore its
collapse to $\ket{\psi^0_c}$ in \eqref{eqn32} cannot by itself indicate a
change in some property of particle $b$. To discuss whether a measurement by
Alice has a physical effect upon Bob's particle requires the use of a framework
in which properties of the latter make sense, both before and after Alice's
measurement takes place. This matter was studied in Ch.~23 of \cite{Grff02c}
for the Bohm version of EPR, showing that there is no such nonlocal effect as
long as Alice's measurement apparatus does not directly interact with Bob's
particle. This is a particular instance of a quite general \emph{Principle of
  Einstein Locality}:%
\begin{quote}
Objective properties of isolated individual systems do not change when 
something is done to another non-interacting system.
\end{quote}
Its proof will be found in \cite{Grff11}. Here ``non-interacting'' means that
the two systems have independent dynamics: the unitary time-development
operator  for the combined systems is the tensor product of the individual
time-development operators of the separate systems. Whether or not the systems
are initially in an entangled state is irrelevant; entanglement should never be
thought of as a mechanism by which one system can `influence' another. This
result is hardly surprising given the widespread acceptance of the no-signaling
principle, since if, contrary to Einstein locality, there were a change in some
objective property, that change could be used to convey information, or at
least this is how a physicist would tend to view the matter.%
\footnote{For an alternative perspective by a philosopher, including a very
  clever construction of an influence that carries no information, see Ch.~4 of
  \cite{Mdln11c}.} %

\xb
\section{\hspace*{.2cm} Conclusion \label{sct7}} 
\xa

\xb
\subsection{\ Summary \label{sbct7.1}}
\xa

\xb
\outl{There are NO nonlocal influences. Cannot be detected. No signaling }
\xa

The central conclusion of this paper is the complete absence of nonlocal
influences between quantum systems which are spatially separated and not
interacting with each other: doing something to one system has no effect,
instantaneous or otherwise, upon the other system. 
Experiments show no evidence of such effects, and the ``no-signaling''
principle, widely accepted in discussions of of quantum information, assumes
their absence.
\ca%
Were such effects present it
should be possible to detect them by means of an experiment, and a confirmed
detection would be worth a Nobel Prize. The ``no-signaling'' principle, widely
accepted in discussions of quantum information, assumes the absence of such
influences. 
\cb%
In brief, if physical reality is quantum mechanical, then quantum
nonlocality, in the sense of nonlocal influences, is a myth.

\xb
\outl{Source of nonlocality idea: Wavefn collapse misunderstood }
\xa

Why, then, the widespread assumption, which often seems taken for granted
without any need to defend it, that quantum mechanics is somehow ``nonlocal''
in a way in which classical physics is not? Wavefunction collapse, produced by
measurements when applied to a system in an entangled state with a distant
system, is one source of the nonlocality notion, and this reflects the
inadequate treatment of measurements in textbooks and much of the quantum
foundations literature. As shown in Sec.~\ref{sct6}, wavefunction collapse is
simply a method of computing a conditional probability, as in classical physics
when two particles are statistically correlated. While this method of
calculation might sometimes be useful in terms of intuitive insight, it does
not correspond to a physical process.

\xb
\outl{Main nonlocality source: Claim local world $\Ra$ Bell inequalities hold}
\xa

\xb
\outl{Refuted by local $^{21}$Ne; factorization formula requires Cl HVs}
\xa

\xb
\outl{Bell $\leq$s assume micro world is \emph{classical}}
\xa

The principal source of the current widespread belief in quantum nonlocality is
undoubtedly the claim by Bell and his successors that in a local world certain
statistical correlations must satisfy some type of Bell inequality. The CHSH
inequality, which belongs to this category, was studied in Sec.~\ref{sct2}
where it was shown that it is violated by quantum correlations which have
nothing to do with spatial separation, but are already exhibited by states
associated with the spin $3/2$ ground state of a $^{21}$Ne atom. This was
followed in Sec.~\ref{sct4} with a discussion of the factorization formula
which is central to derivations of Bell inequalities, and makes reference to a
hidden variable or variables, typically denoted by $\lm$. Such hidden variables
are always assumed to be classical; they lack the structure of noncommuting
projectors which are central to Hilbert space quantum mechanics. It is
regrettable that so much attention has been paid to the locality assumptions
involved in the derivation of Bell inequalities, and so little to the equally or
more important assumption that quantum probabilities can be discussed using a
classical sample space: in essence, assuming the microscopic world is not
quantum mechanical but classical.

\xb
\outl{Qm measurement not properly discussed in textbooks. }
\xa

\xb
\outl{(First) measurement problem solved by choice of pointer PDI}
\xa

\xb
\outl{Second problem solved using histories. Single framework rule is needed }
\xa

Much of the confusion surrounding discussions of nonlocality has to do with the
absence from standard quantum mechanics, understood as what is found in
textbooks, of a proper discussion of quantum \emph{measurements}, and for this
reason the essential principles have been summarized in Sec.~\ref{sct3}. The
key to resolving what is generally referred to as the \emph{measurement
  problem}, the possible appearance of superpositions of macroscopic `pointer'
states (Schr\"odinger cats), is to use the consistent histories formulation of
quantum theory in which time development is represented by stochastic histories
rather than restricted to the unitary time development of a wavefunction. Using
a \emph{framework} (family of histories) with projectors for the pointer states
gets rid of this measurement problem. Using a framework in which these
macroscopic measurement outcomes are correlated with microscopic properties of
the measured system at a time just before the measurement took place, resolves
a second measurement problem: how the macroscopic outcomes can be used to infer
(retrodict) the prior microscopic property that resulted in (caused) a
particular outcome. Consistent reasoning using frameworks requires paying
attention to the (possible) noncommutativity of quantum projectors as embodied
in the \emph{single framework rule}.

\xb
\outl{Bell $\leq$s error: Factorization condition assumes single Cl sample 
space}
\xa

The tools used to analyze measurements in a fully quantum mechanical fashion
made it possible to identify, in Sec.~\ref{sct4}, the fundamental error, from
the perspective of a consistent quantum theory, in derivations of the CHSH and
other Bell inequalities. It is the assumption that the factorization condition
\eqref{eqn24} for probabilities can use \emph{classical} hidden variables
(parametrized by the symbol $\lm$) associated with a \emph{single} sample
space, rather than appropriate quantum sample spaces, projective decompositions
of the identity (PDIs).

\xb
\outl{EPR: did not understand measurements; false counterfactual assumption}
\xa

\xb
\outl{Einstein was right: there are no ghostly nonlocal influences}
\xa

Bell's work was motivated by the Einstein-Podolsky-Rosen (EPR) paper, in which
locality was simply assumed, and the claim was made that quantum mechanics is
incomplete. Their work was based on an inadequate understanding of quantum
measurements, which at that time were assumed to simply collapse wavefunctions.
In addition, their argument employs a counterfactual assumption which,
translated into the Bohm version of the EPR paradox, is that while Alice
actually measured (say) $S_z$, she could instead have measured and obtained a
value for $S_x$ during this particular run. But if one assumes, contrary to
EPR, that Hilbert space quantum mechanics is complete, such a counterfactual
assumption is misleading, since a spin-half particle cannot simultaneously
possess an $S_x$ and an $S_z$ property. That has nothing to do, at least in any
direct sense, with the EPR locality assumption. On the other hand, Einstein's
belief that there are no ghostly nonlocal influences (``spukhafte
Fernwirkungen'') is fully justified, as noted in Sec.~\ref{sct6}, by a
consistent analysis employing Hilbert subspaces resulting in a Principle of
Einstein Locality. 

\xb
\outl{EPR correlations explained by \emph{quantum} common cause}
\xa

An additional argument, Sec.~\ref{sbct5.3}, undermines claims for quantum
nonlocality based on correlations that violate Bell inequalities by showing that
the relevant \emph{quantum} correlations can be understood as arising from a
\emph{local quantum common cause}, something which, in the case of the
polarization of down-converted photons, occurs at the source where they were
created. This understanding makes use of the analysis of quantum measurements
in Sec.~\ref{sct3}, in particular the fact that measurement outcomes 
reflect earlier microscopic properties of the measured system when analyzed
using an appropriate framework.

\xb
\subsection{\ Terminology: Some Suggestions \label{sbct7.2}}
\xa

\xb
\outl{Use of misleading terminology. Example of `heat capacity'}
\xa

Even the reader who agrees with the arguments presented in this paper may
nonetheless, and with some justification, take the attitude that scientific
terminology often acquires a technical meaning that is different from the way in
which it was first used, and hence there is no difficulty if `local' and 
`nonlocal' continue to be used in the same way as in much of the current
literature on quantum foundations and quantum information. After all, there
are other examples: the term `heat capacity' is in common use in
thermodynamics, and no one, except perhaps beginning students, is confused by
the fact that `heat' is no longer regarded as a fluid, and heat capacities are
typically measured by doing work on the system of interest, rather than
connecting it to a thermal reservoir.

\xb
\outl{'Heat capacity' sometimes OK; Qm `nonlocality' almost always wrong}
\xa

However, in the case of `heat capacity' there are at least some circumstances
in which heat can, indeed, be treated as a conserved fluid, whereas in quantum
mechanics `nonlocality' seems in almost every respect a misleading and
confusing term. Granted, students who are setting up apparatus in the
laboratory are, at least after a while, not likely to worry that an experiment
set up at some distant location might suddenly make a photon disappear while on
its way through an optical fiber to a detector, or perhaps suddenly appear out
of nowhere. Theoreticians are more likely to be confused by nonlocality claims,
and the appearance of such claims in textbooks and the popular literature can
only add to the confusion felt by students learning quantum theory for
the first time.

\xb
\outl{Suggest: BELL nonlocal; SCHRODINGER steering; ``Qm world is LOCAL'' }
\xa

Those who agree with the author that clear thinking is a key part of good
physics, and using appropriate terms is an aid to clear thinking, might at
least wish to consider some alterations and/or clarifications in the use of
various terms. Replacing `nonlocal' with `Bell nonlocal', a term already used
in some publications, would be a useful clarification, and certainly
appropriate, in that Bell himself believed (incorrectly) that violations of his
inequalities indicated nonlocality. Similarly, replacing `steering' with
`Schr\"odinger steering' would be a step in the right direction. However, in
both cases adding a comment that the quantum world is in reality local---there
are no instantaneous long-range influences---would help counter a widespread,
but mistaken, belief to the contrary.

\xb
\outl{Replace or supplement `local' with `classical', in `local
  realism/causality'}
\xa

Replacing, or at least supplementing, `local' with `classical' in certain
phrases would also be an improvement. Thus claims, e.g. \cite{Hnao15ea,Wsmn15},
that recent experiments show that quantum mechanics is inconsistent with `local
realism' lead to the strange conclusion that if quantum mechanics is local (as
argued here) it must be unreal. But we have ample evidence that the real world
is best described by quantum, not classical, mechanics, and so it is
`\emph{classical} realism' that is ruled out by experiments. Similarly,
replacing `local causality' as used in \cite{Bll90d,Nrsn11} with
`\emph{classical} local causality' as a key ingredient in the derivation of
Bell inequalities would help clarify their true nature.

These are simply offered as suggestions; the author does not wish to lose
friends (assuming he still has some) through disputes over terminology. The
goal should be to use terms, including technical terms, which aid clear
thinking rather than creating confusion.

\xb

\end{document}